\documentclass{elsart}
\usepackage{graphicx}
\usepackage{natbib}
\usepackage{amssymb}

\begin{document}

\begin{frontmatter}
\title{Correlation between Risk Aversion and Wealth distribution}

\author[UFRGS]{J. R. Iglesias\corauthref{cor}}
\corauth[cor]{Corresponding author} \ead{iglesias@if.ufrgs.br}
\author[UFRGS]{ S. Gon\c{c}alves}
\author[CAB]{G. Abramson}
\author[BVBA]{J.L. Vega}

\address[UFRGS]{Instituto de F\'{\i}sica, Universidade Federal do Rio Grande
do Sul, C.P. 15051, 91501-970
Porto Alegre RS, Brazil}
\address[CAB]{Centro At\'{o}mico Bariloche, Instituto Balseiro and CONICET,
8400 San Carlos de Bariloche, Argentina}
\address[BVBA]{Banco Bilbao Vizcaya Argentaria, Via de los Poblados s/n, 28033
Madrid, Spain}

\thanks{J.R.I. acknowledges support from CNPq (Brazil) and the
hospitality and support of Laboratoire Louis N\'eel, CNRS,
Grenoble and Laboratoire de Physique des Solides, Universit\'{e}
Paris-Sud, Orsay, France. S. G. and J.R.I. thank the hospitality
of Instituto Balseiro and Centro At\'omico Bariloche, S.C. de
Bariloche, Argentina, and G.A. thanks the hospitality of the
Instituto de F\'{\i}sica, Universidade Federal do Rio Grande do
Sul, Porto Alegre, Brazil. We acknowledge partial support from
SETCYP (Argentina) and CAPES (Brazil) through the
Argentine-Brazilian Cooperation Agreement BR 18/00.}

\begin{abstract}
% Text of abstract
Different models of capital exchange among economic agents have
been proposed recently trying to explain the emergence of Pareto's
wealth power law distribution. One important factor to be
considered is the existence of risk aversion. In this paper we
study a model where agents posses different levels of risk
aversion, going from uniform to a random distribution. In all
cases the risk aversion level for a given agent is constant during
the simulation. While for a uniform and constant risk aversion the
system self-organizes in a distribution that goes from an unfair
``one takes all'' distribution to a Gaussian one, a random risk
aversion can produce distributions going from exponential to
log-normal and power-law. Besides, interesting correlations
between wealth and risk aversion are found.
\end{abstract}

\end{frontmatter}

Probably one of the most important contribution to the study of
the distribution of personal income and wealth was made at the end
of the XIXth century by Italian economist Vilfredo Pareto. In his
book ``Cours d'Economie Politique''\cite{Pareto}, he presented the
statistical analysis of the income distribution of different
European regions and countries. He concluded that the income
distribution follows a rather universal law, characterized by a
logarithmic pattern, described by the formula: $logN \varpropto
\alpha log(w)$; where $N$ is the number of income earners with
income higher than $w$ and the exponent $\alpha$ is named Pareto
index. This income distribution is a power law and Pareto
determined the exponent to be $1.2 \leq \alpha \leq 1.9$. Pareto's
law is a classical example of a distribution with no
characteristic scale and, as asserted by Pareto himself, a low
value of the slope $\alpha$ denotes a higher inequality of the
income distribution.

\begin{figure}
\centering
\includegraphics[angle=0, width=10cm]{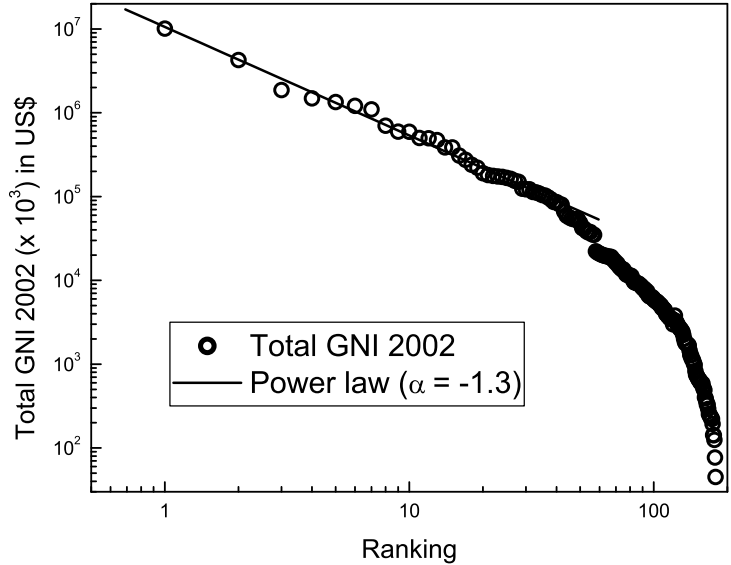}
\caption{GNI ranking of 179 countries\cite{WB}}\label{gni2002}
\end{figure}

However, analysis of current economic data seems to indicate that
Pareto's law is valid for the high income strata of society, while
for middle and low income classes the distribution appears to be a
log-normal (Gibrat) distribution. Data for
Japan\cite{souma,fuji,aoyama}, the United Kingdom and the United
States of America\cite{dragu2000,dragu2001} confirm this idea.
Also, in Fig. \ref{gni2002} we present a plot in log-log scale of
GNI (Gross National Income) data of 179 countries \cite{WB} where
it is possible to make a power law fit for the richer countries,
while the others seems to be better fitted by an exponential or
log-normal distribution.

A great deal of effort has been devoted to obtain the power law
distribution of the wealthiest
strata\cite{Matteo,Chakra,west,ispo98,Bouchaud,solo}. In previous
articles we have proposed a Conservative Exchange Market Model
(CEMM)\cite{PIAV2003,IGPVA2003} with extremal dynamics of the kind
of Self-Organized Criticality (SOC) theories\cite{SOC1,SOC2}. The
obtained distribution was a Gibbs-exponential type and the results
were in good agreement with the distribution of some welfare
states such as Sweden\cite{SI2004}. Other
authors\cite{Chakra,west} have proposed models in which agents
save a fraction of their capital, and only the rest may be
exchanged. In the language of economics this \emph{saved part of
the resources} is a measure of the agent risk aversion. Following
these ideas, we present here a family of models that combine the
risk aversion ingredients with Monte Carlo dynamics and extremal
dynamics. We found different interesting shapes for the wealth
distribution, and in some particular cases a power law profile is
obtained.

Let us consider a set of economic agents characterized by a risk
aversion factor $\beta(i)$, so that $[1-\beta(i)]$ is the
percentage of wealth the $i-$agent is willing to risk. An agent
with $\beta(i) = 0$ is a radical one who risks all his assets,
while, on the other hand, $\beta(i)=1$ characterizes a totally
conservative agent who simply does not play the game. The dynamics
of the system is as follows: For the choice of the two partners,
one can adopt an extremal dynamics as in previous
calculations\cite{PIAV2003,IGPVA2003}, or a Monte Carlo method as
in refs. \cite{Chakra,west,ispo98}. In the first case we start by
determining the site with the minimum wealth, and then we choose
the other partner of the exchange at random. In the second case
both agents are chosen at random, independently of their wealth.
When considering the case of extremal -minimum - dynamics we model
the situation where the poorest agent will try to do something to
improve its situation, or else, some external regulator
(government, for example) will act in order to favor the
handicapped. In that case one expects a more equitable wealth
distribution. The second case is best adapted to represent a kind
of stock market, where the transactions occur independently of the
fortune of the agents, and in general the poverty line is strictly
zero, that is agents can bankrupt. In both cases, we prescribe
that no agent can win more than he puts at stake, so the value
that will be exchange is the minimum value of the available
resources of each agent, i.e.
$dw=min[(1-\beta_1)w_1;(1-\beta_2)w_2]$.

Finally, we introduce a probability $p \geq 0.5$ of favoring the
poorer of the two partners, because {\it a stable society requires
that the poor have an advantage in transactions with the wealthy
and are protected by particular rights and marketing
freedom}\cite{west}. Increasing the probability of favoring the
poorer agent is a way to simulate the action of the state or of
some type of regulatory policy that tries to redistribute the
resources\cite{IGPVA2003,SI2004}. We consider two cases: a) a
fixed probability $p$ going from 0.5 to 1 and b) a random value of
$p$, making use of the expression proposed in ref \cite{west},

\begin{equation}
\label{eq:sca} p=1/2+f\times\frac{w_1-w_2}{w_1+w_2}
\end{equation}

\noindent  $w_1$ being the wealth of the richer partner and $w_2$
that of the poorer one; $f$ is a factor going from $0$ (equal
probability for each agent) to $1/2$.

We consider the number of agents $N$ ranging from $10^4$ to $10^6$
and a number of average exchanges going from $10^3$ to $10^4$ per
agent. In addition to the two different types of dynamics for the
system we present here results for a) $\beta$ and $p$ uniform b)
$\beta$ and $p$ random. The discussion of the first case, although
rather idealized, is important in order to have a clear idea of
the effect of risk aversion and of the probability of having or
not a better treatment for the disfavored layers of the
population. The second case is a more realistic vision of an
heterogeneous society. Of course, in the real world, agents can
change their risk strategy as a function of the results. This
possibility will be discussed in a forthcoming article.

a) {\it Uniform $\beta$ and $p$}

\begin{figure}
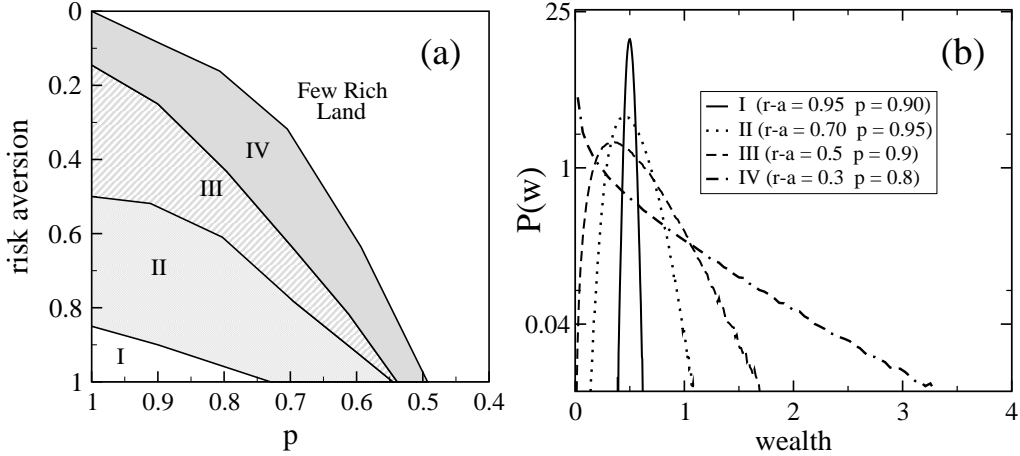
 % [H]
\includegraphics[height=6cm, clip=true]{modII_diag.eps}
\includegraphics[height=6cm, clip=true]{modII_dist.eps}
\caption{Model with uniform $\beta$ and $p$, for $N=10^5$ and
$10^3$ transaction per agent. Region I correspond to a very narrow
wealth distribution, {\it utopian socialism}, region II and III
present skewed Gaussians and region IV corresponds to an
exponential distribution. Outside these regions there is no a true
wealth distribution because in the {\it few rich land} one or few
agents concentrate all the available resources while the others
have strictly zero wealth} \label{modII_dist}
\end{figure}

In this rather hypothetical situation all agents have the same
risk aversion parameter $\beta$, and all transactions have the
same probability $p$ of favoring the poorer agent. Both parameters
are constant during the simulation. Let´s first present the
results for the Monte Carlo simulation, where both agents are
chosen at random. The results are summarized in the diagram
depicted in Fig.~\ref{modII_dist} The different regions correspond
to different types of resulting wealth distributions. Wealth
distributions in Region I are very narrow and Gaussian-like, so we
call this region {\it Utopian socialism} because almost all agents
have the same income with a small dispersion. Region II has
Gaussian-like distributions too but skewed to higher values of
wealth, therefore we named it {\it liberal socialism}. Next region
(III) has hybrid wealth distribution, Gaussian-like for low wealth
values, and exponential for high wealth values, and we call it
{\it moderated capitalism}. In the last region (IV) wealth
distribution are exponentials with a tendency to power-laws, so we
call this region of {\it ruthless capitalism}. The different type
of wealth distribution in the four regions can be seen in
Fig.~\ref{modII_dist} for some typical values of the parameters
$\beta$ and $p$. Outside Region IV there is a region of parameters
that we called the {\it few rich land} as the outcome of the
dynamics ultimately favors just one (or a very few) agent which
concentrates all the available resources. In this later case,
since no more exchanges are possible, the system freezes: a very
greed economy carries in itself its own destruction. Obviously the
$\beta=1$ line is of no interest, and the same is true for the $p
\leq 0.5$ region which is always in the few rich land.

Simple as it is, this model captures the essence of general
economic exchanges, considering the resulting wealth distribution
corresponding to different economic policies old and present. It
is amusing that just playing with the two numerical parameters of
the model very different behaviors are obtained. {\it Utopian
socialism}, for example, exhibits slight economical differences
between agents and this is due to the combined force of high
values of $\beta$ and $p$, which means a repressed market and a
strong intervention favoring the poorer. The gradual
liberalization of the market, through lower risk aversion (less
controlled market) and less state intervention in the social sense
(lower values of $p$), gives rise to more liberal economies with
higher inequalities in the income.

We have also performed simulations for this case, but using
extremal dynamics of the type described in
refs.\cite{PIAV2003,IGPVA2003}. That means that one of the
partners is the agent with the minimum wealth, while the second
one is chosen at random. The results are rather odd. For low
values of $p$, the dynamics of the system freezes with no
subsequent economic activity, because the agent with minimum
wealth has no resources to exchange, so the system converges to
zero activity. One possible way to overcome this situation should
be to consider a different asset transfer rule. On the other hand,
for $0.7 \leq p \leq 1.$ the minimum dynamics generates an
exponential distribution, where almost all the agents lie in the
middle class. However, for some values of $\beta \geq 0.7$ and $p
\approx 1$ the middle class is split in two income regions with a
gap in between. This is probably so because of some kind of
anti-resonance combined effect of the rules of the dynamics and
the conservation constraint.

b) {\it Random $\beta$ and $p$}

In a more realistic approach to the risk preferences in the
population, we consider a disordered risk aversion parameter
throughout the system. Each agent is assigned a value of
$\beta_i$, drawn at random from a uniform distribution on the
interval $(0,1)$. We consider only quenched disorder, where each
agent maintains his risk aversion despite the outcome of the
exchange, disregarding his own success or failure in increasing
her wealth. Simulations have been carried out for different values
of the probability $p$ and also for the complete range of the
asymmetry parameter $f$ in Eq.\ref{eq:sca}.

\begin{figure}[ht]
\includegraphics[angle=270, width=6.7cm]{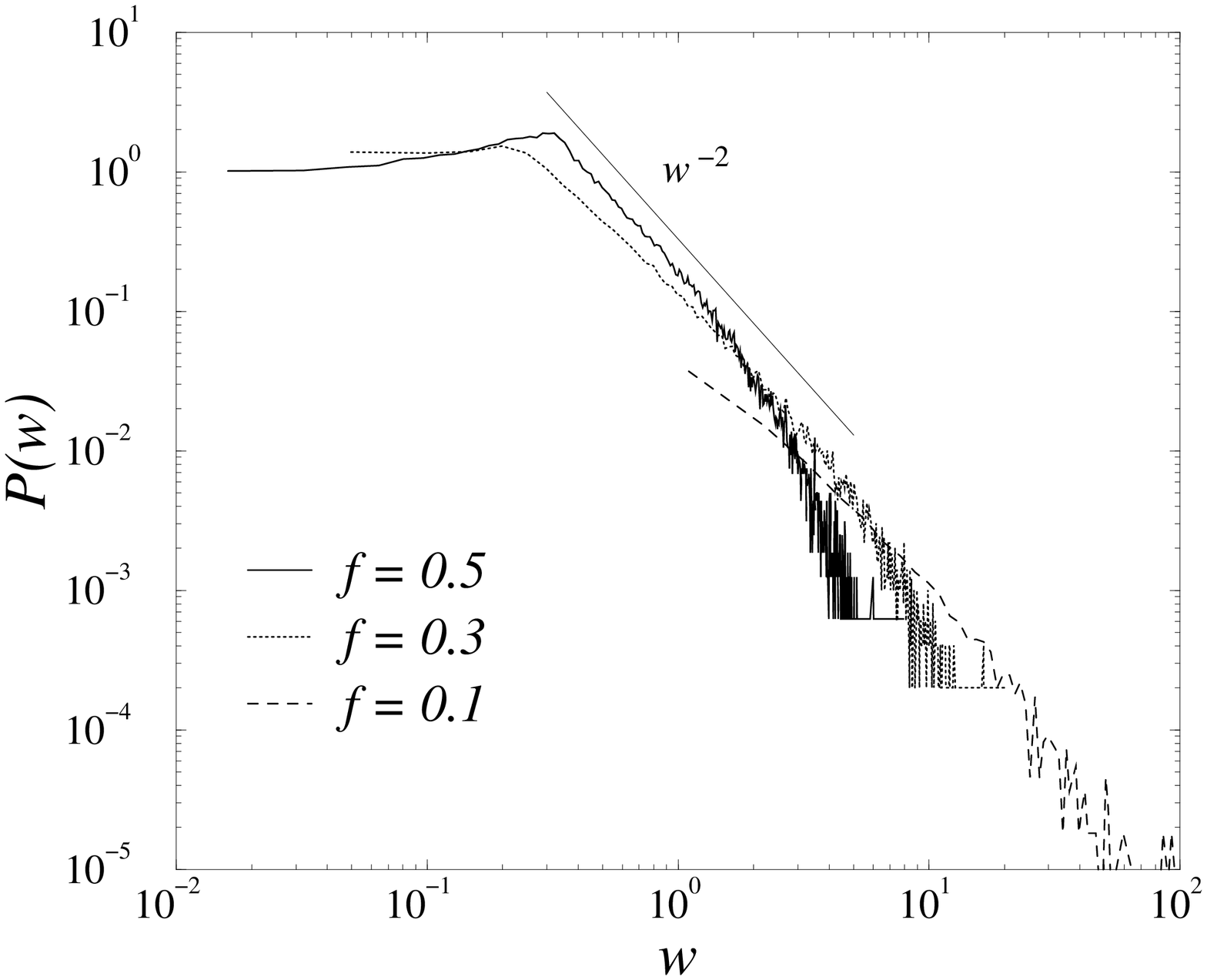}
\includegraphics[angle=270, width=6.7cm]{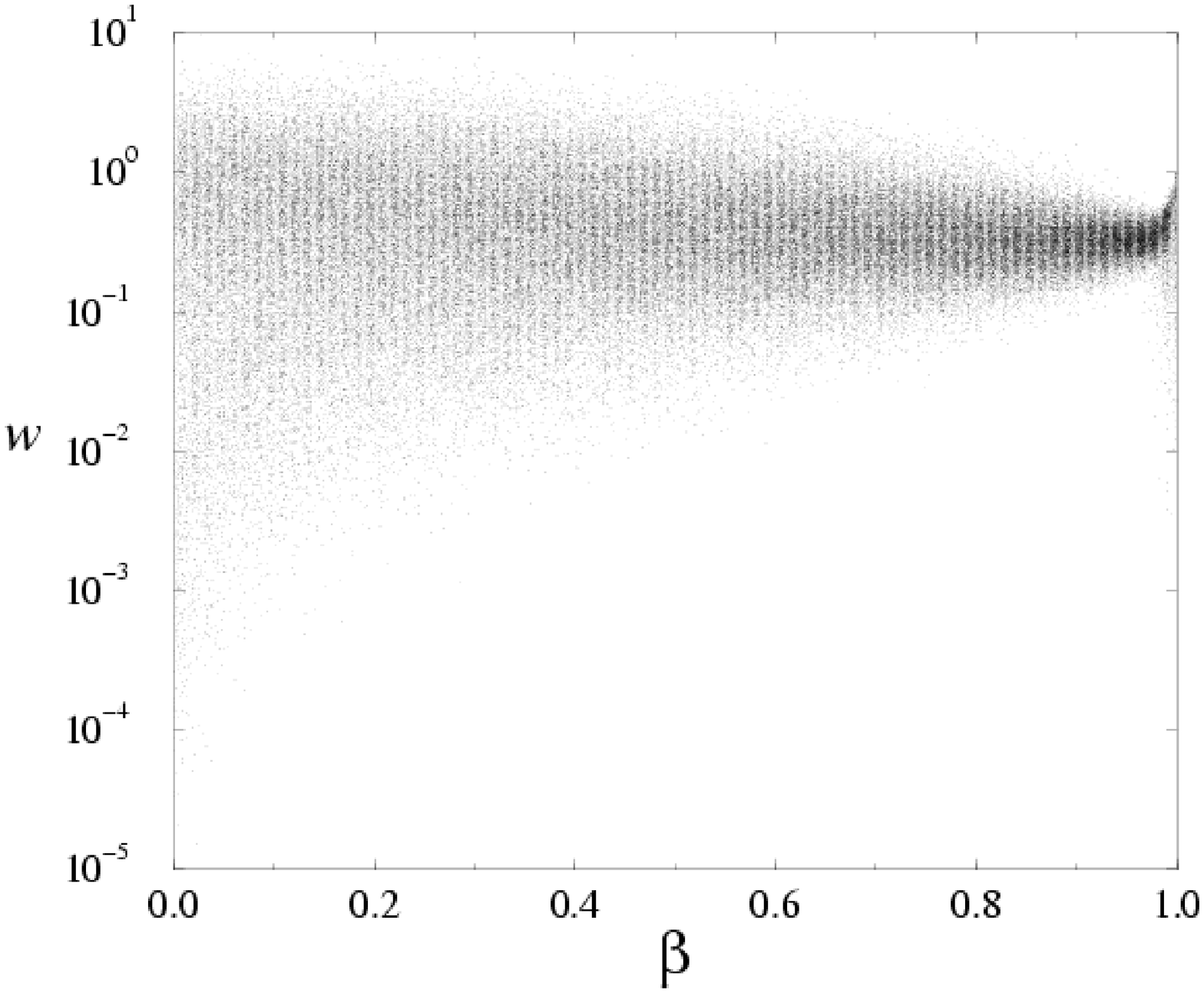}
\caption{Left panel: Wealth distribution for random $\beta$ and
Monte Carlo dynamics, the distribution is calculated for $N=10^5$
and $10^4$ exchanges per agent on average. Results are shown for
three values of the asymmetry parameter, as shown in the legend. A
$x^{-2}$ curve is also shown to guide the eye. Right panel:
Correlation between wealth and saving parameter for $N=10^5$,
$f=0.5$} \label{betarandom}
\end{figure}

Some typical distribution curves are shown in the left panel of
Fig.~\ref{betarandom}. For $f=0$, i. e. when trades do not favor
either of the partners, the distribution of wealth becomes, slowly
but steadily, a delta function at $w=0$, with the wealth
concentrated in one or a few agents, and the rest owning
effectively nothing (this results are not shown in
Fig.~\ref{betarandom}). This can happen, even though each agent
risks only part of his capital at each interaction, because there
is no restriction in the amount he can loose in successive
exchanges. The situation is a multiplicative process with $w=0$ as
an absorbing point. On the contrary, when the externally imposed
asymmetry favors, statistically, the poorer agent, we observe the
emergence of a distribution characterized by three regimes. There
is a peak in the distribution that separates a poor class to the
left, from a middle class to the right (see for example the full
line in Fig. \ref{betarandom}, corresponding to $f=0.5$). The
middle class follows a power law distribution of the form
$P(w)\sim w^{-\alpha}$, with $\alpha$ depending on the value of
$f$, and $\alpha \approx 2$ for $f=0.5$. Finally, there is a
transition from this power law behavior to an exponential tail
encompassing the wealthier agents. This exponential tail is not an
effect of the finite size of the system, as has been verified for
system sizes up to $10^6$. Also one can observe that there is a
finite number of very poor people, contrasting with previous case
(a). We have also represented on the right panel of Fig.
\ref{betarandom} the correlation between wealth and risk aversion.
One observes that the range of wealth variation is up to five
magnitude orders. Besides that, on average, the higher values of
income correspond to a high risk aversion, while the highest
individual wealths correspond to risk-loving agents. But also the
lowest incomes belong to risky agents, as expected.

\begin{figure}[ht]
\includegraphics[angle=0, width=7cm]{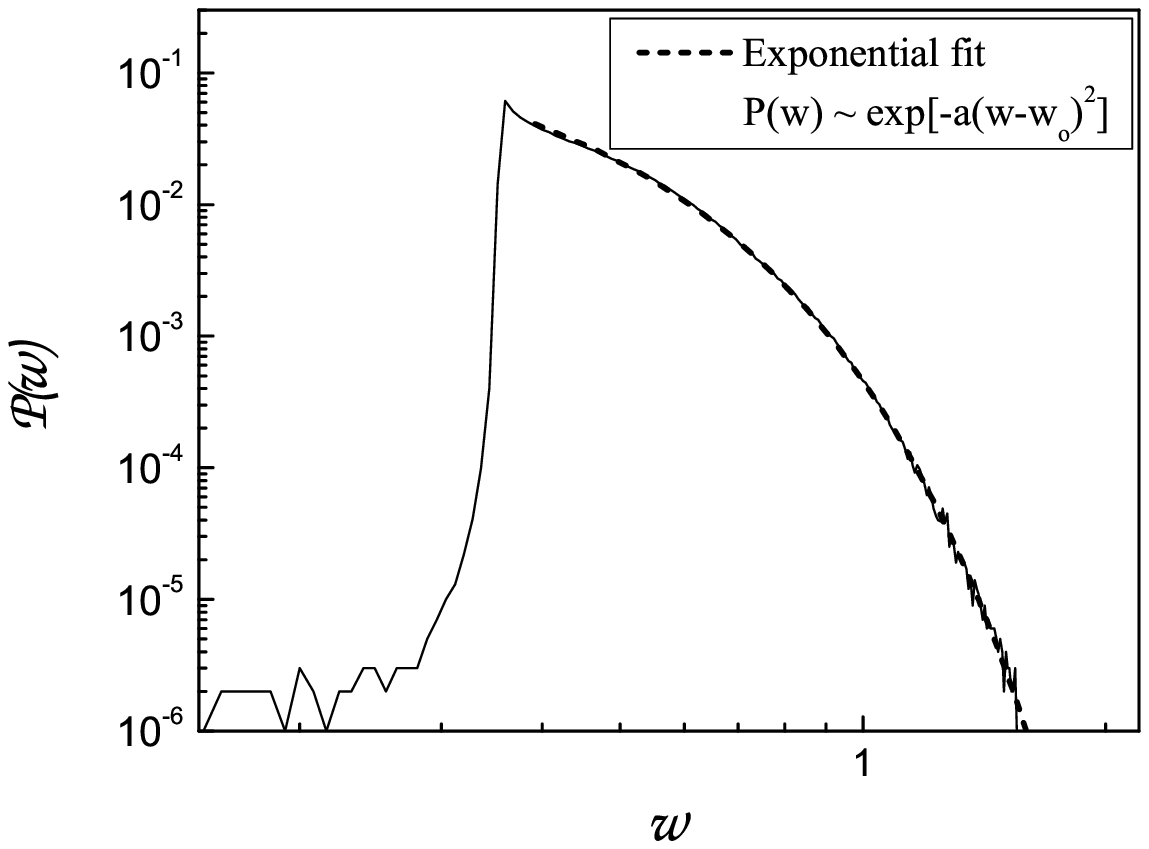}
\includegraphics[angle=0, width=6.8cm]{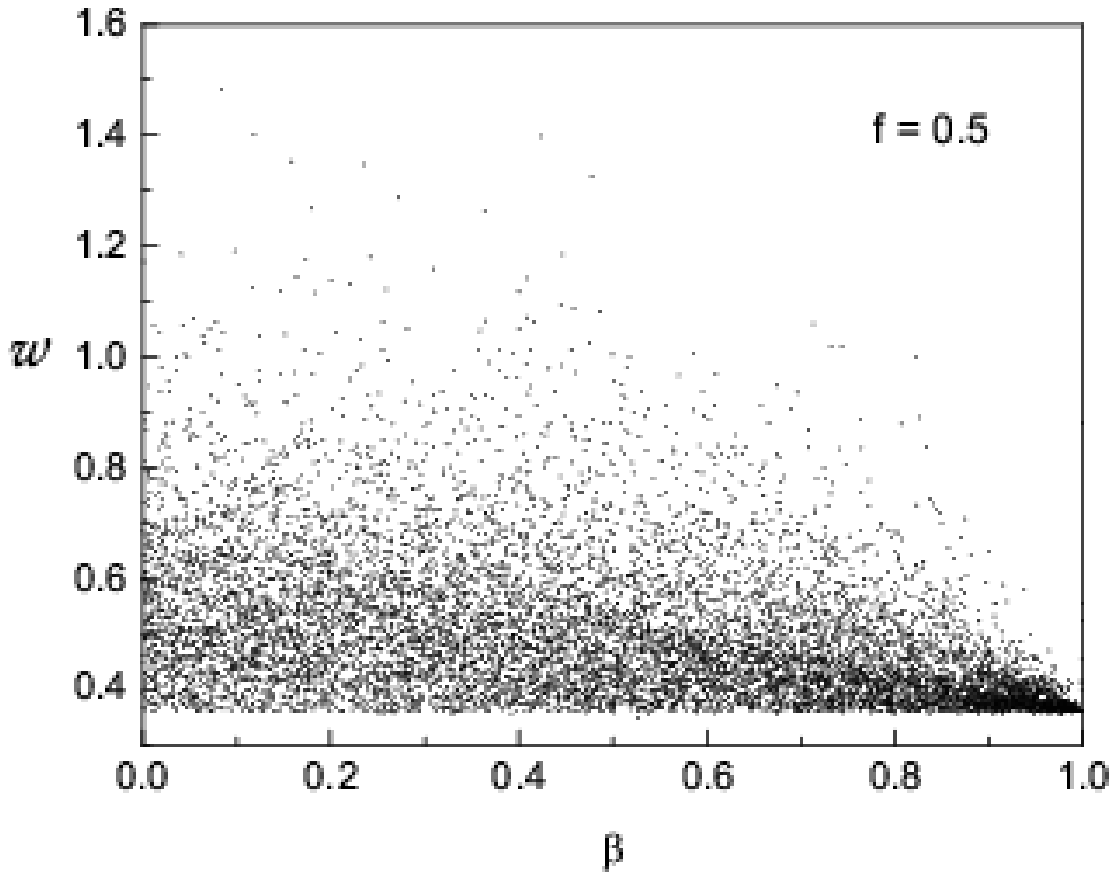}
\caption{Left panel: Wealth distribution for random $\beta$ and
minimum dynamics for $N=10^5$ and $10^4$ exchanges per agent, in
average, and for $f=0.5$. The high income region is well fitted by
a kind of exponential distribution with $a \approx 1.1$ and $w_o
\approx 0.7$. Right panel: Correlation between wealth and saving
parameter for the same values of $N$ and $f$. } \label{betasoc}
\end{figure}

When considering extremal (minimum) dynamics, the results are
quite different as they are shown on Fig.\ref{betasoc}. It is
possible to see that it appears a threshold, or {\it poverty
line}, and also that the distribution is very narrow compared to
the Monte Carlo case. One observes that there are vary many few
people below the poverty line and its value is around $0.38$ for
$f = 0.5$
Moreover, the high income region behaves in an exponential way,
following a law of the type $P(w) \approx exp\{-a(w-w_o)^2\}$ with
$a \approx 1.1$ and $w_o \approx 0.7$ for $f = 0.5$. The poverty
line is well seen too on the right panel of Fig.\ref{betasoc}, all
the agents are above $0.38$ and now it is clear that a low risk
aversion favors, on average, a higher wealth.

To conclude: taking into account risk aversion (or saving, as
defined by others authors\cite{Chakra,west}) generates a rich
variety of wealth distributions, when combined with different
choices of trading rules. For some particular values of the
exchange probability $p$ and a random choice of $\beta$ a power
law profile is obtained. Here we have compared in detail an
extremal and Monte Carlo dynamics for constant and random risk
aversion and a simple exchange bias. Extremal (minimum) dynamics
provides a more equitable society, in the sense proposed in the
classical work by J. Rawls\cite{Rawls}: {\it no redistribution of
resources within... a state can occur unless it benefits the least
well-off}, and this is clear because of the existence of a poverty
line and the emergence of a wealth distribution with a large
middle class. Monte Carlo dynamics seems to better reproduce a
capitalist society: there are very many people with almost zero
income and one can observe a power law distribution for the higher
layers of the social spectra. Of course people can change its risk
strategy as a function of their own results. We have investigated
this possibility and the results will be published elsewhere.

\end{document}